\documentclass[10pt]{iopart}
\usepackage{graphicx}

\newcommand{\sign }{\textrm{sign}}

\begin{document}
\jl{1}
\title{Counting nodal domains}
\author{Georg Foltin}
\address{Institut f\"ur Theoretische
Physik III, Heinrich--Heine--Universit\"at D\"usseldorf,
Universit\"atsstrasse
1, D--40225 D\"usseldorf, Germany}
\begin{abstract}
We consider the nodal domains of Gaussian random waves in two dimensions. We present a method to calculate the distribution of the number of nodal domains and the average connectivity with the help of auxiliary Potts-spins. An analytical approach could   be helpful to decide whether the pattern of nodal domains belongs to the universality class of short-ranged percolation. This is not completely evident since we find (weak) long-ranged correlations between distant avoided nodal intersections.
\end{abstract}

\pacs{05.45, 02.50, 03.65}

\section{Introduction}
In the classical world one can easily distinguish between integrable and chaotic systems just by looking at the system's trajectories. Upon increasing $\hbar$ and entering the quantum world, the notion of trajectory loses its meaning and more subtle measures to characterize quantum chaos are needed. The most common method to reveal the quantum chaotic nature of a system is to look at the spectral statistics of its Hamiltonian. The spacings of the spectral lines have a Wigner-Dyson distribution instead of a Poisson or regular distribution. Recently, Blum et. al. \cite{Blu02} proposed an alternative measure based on the distribution of the number of nodal domains of the \textit{wave function}. To have a specific system in mind we consider a particle in a (two-dimensional) quantum billiard. In the absence of a magnetic field one can choose real wave functions. The nodal domains are now the regions of constant sign; they are bounded by the nodal lines where the wave function vanishes. 

\begin{figure}[b]
\includegraphics[scale=0.5]{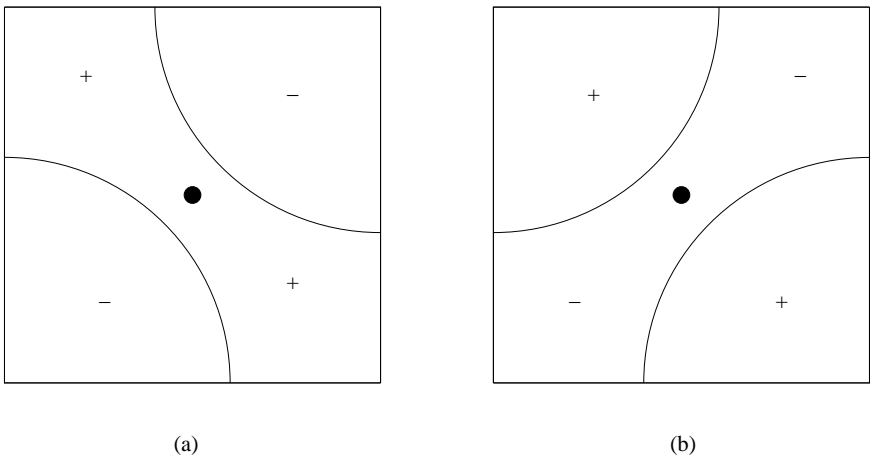}
\caption{\label{bog}Both types of avoided intersections in the model \cite{Bog02}}
\end{figure}

In case of an separable system, one observes a grid of intersecting nodal lines, whereas in the chaotic case  the nodal domains form a highly irregular pattern and nodal intersections are rare \cite{Str79,Sim96}. Similar pattern are also found in Gaussian random waves, which are known to be a good model for chaotic wave functions \cite{Ber77}. Here, nodal intersections hardly ever occur, i.e.  saddle points and nodal lines  coincide with vanishing probability. Instead, depending on the sign of the wave function right at a saddle point, junctions between positive (negative) regions are formed \cite{Str79,Pec72,Mon02} (avoided intersections), as illustrated in figure \ref{bog}. Bogomolny and Schmit \cite{Bog02} mimic the distribution of nodal domains by the following procedure. They start from a checkerboard-like lattice of positive and negative regions, where the (linear) size of the squares corresponds to the wavelength of the random waves. At each intersection they place with equal probability and independently a junction (see figure \ref{bog}) which either connects the positive or negative regions. Since their model belongs to the universality class of critical, \textit{short-ranged} percolation, they are able to compute the moments of the numbers of nodal domains, the area distribution and corresponding (critical) exponents.

In this article we argue, that the applicability of short-ranged percolation to the nodal domains of Gaussian random waves is not entirely obvious. We will demonstrate, that distant avoided intersections are in fact correlated. Since the correlations are rather weak, they will if at all modify the critical properties of very large systems, which cannot be accessed numerically, calling for  an analytical scheme to compute the number of domains and their (mean) connectivity. We put forward a possible method in the second part of the article. We hope that it will ultimately lead to an analytical approach to the problem of counting nodal domains. 

\section{Long-ranged correlations of avoided intersections in random waves}
Gaussian random waves might be seen as a superposition $\phi=\sum_\bi{k}a_\bi{k}\phi_\bi{k}$ of partial waves $\phi_k(\bi{r})$ with fixed energy $k^2$ and amplitudes $a_\bi{k}$ with a Gaussian distribution. The random waves obey the Helmholtz equation $-\nabla^2\phi=k^2\phi$ and are characterized (in two dimensions) by the correlation function \cite{Blu02,Ber77,Bog02}
\begin{equation}\label{corr}
\left<\phi(\bi{r})\phi(0)\right>=J_0(r)\sim \cos(2\pi r)/\sqrt{r} \textrm{ for }r\rightarrow\infty
\end{equation}where we have set the energy to $E=k^2=1$. The Gaussian average is denoted by $\left<\ldots\right>$.
The nodal domains are the (connected) regions of constant sign of the random field $\sigma(\bi{r})=\sign(\phi(\bi{r}))$, the nodal lines are at $\phi\equiv0$.

In \cite{Bog02} the pattern of nodal domains is modelled by a staggered (checkerboard-like) lattice of positive and negative regions. With equal probability (and at each node independently)  either  two positive or two negative areas are connected (see figure \ref{bog}).
The equivalent quantity for the random waves is simply the sign of the random field $\phi$ right at the saddle points---if the sign of $\phi$ is positive at the saddle point, the positive regions are connected and vice versa.

Therefore, we have to calculate correlations of the density
\begin{equation}
\chi=-\sign(\phi) \theta(-\det(\partial_i\partial_j\phi))\det(\partial_i\partial_j\phi)\delta^2(\nabla \phi).
\end{equation}
The $\delta$-function picks up the extrema of $\phi$ (where $\nabla\phi=0$), the determinant of $\partial_i\partial_j\phi$ yields the corresponding normalization, and the theta-function singles out the saddle points, which have a negative determinant, see e.g. \cite{Mon02}.

The calculation of $\left<\chi(\bi{r})\chi(0)\right>$ is done in an expansion in terms of $J_0(r)$. One might think of $\chi$ as a series in $\phi$ and its derivatives (to obtain the series expansion, one has to replace at first the $\delta$, $\theta$, and sign-function by their respective smooth counterparts). Now we perform the Gaussian average with the help of the Wick theorem. We have \textit{internal} contractions within $\chi(\bi{r})$ and within $\chi(0)$, and contractions \textit{connecting} $\chi(\bi{r})$ and $\chi(0)$. We sort the expansion by the number of the latter connections. The first  term of the expansion is obviously
\begin{equation}
\left<\chi(\bi{r})\chi(0)\right>=\int\rmd^2r'\int\rmd^2r''\left<\frac{\delta\chi(\bi{r})}{\delta\phi(\bi{r}')}\right>J_0(|\bi{r}'-\bi{r}''|)
\left<\frac{\delta\chi(0)}{\delta\phi(\bi{r}'')}\right>+\ldots
\end{equation}
It contains all terms, which decay as $J_0(r)\sim r^{-1/2}$, whereas the next term in the expansion decays not slower than $(J_0(r))^3\sim r^{-3/2}$ (since $\chi$ is antisymmetric in $\phi$ the term $\propto (J_0(r))^2$ vanishes).

To obtain the functional derivative of $\chi$, we add to $\phi$ an infinitesimal (fixed) variation $\phi\rightarrow\phi+\eta$ and calculate $\left<\chi([\phi+\eta],\bi{r})\right>$ to linear order in $\eta$. Since $\partial\phi$ is independent from $\phi$ and from $\partial\partial\phi$ as well, we have
\begin{equation}
\fl\left<\chi([\phi+\eta],\bi{r})\right>=\left< \sign(\phi+\eta) F(-\det(\partial_i\partial_j\phi+\partial_i\partial_j\eta))\right>\left<\delta^2(\nabla \phi+\nabla\eta)\right>,
\end{equation}
where $F(x)=x\theta(x)$. Due to its antisymmetry with respect to $\phi\rightarrow-\phi$ the left expectation value vanishes for $\eta=0$. Therefore, using the derivatives sign$'(x)=2\delta(x)$ and $F'(x)=\theta(x)$
\begin{eqnarray}
\label{varrhoj}
\delta_\eta\left<\chi(\bi{r})\right>&=&\left<\delta^2(\nabla \phi)\right>2\eta\left<\delta(\phi)F(
-\det\partial\partial\phi)\right>\nonumber\\
&&{}-\left<\delta^2(\nabla \phi)\right>\left<\sign(\phi)\theta(-\det\partial\partial\phi)\partial_i\partial_k\phi\right>
\epsilon_{ij}\epsilon_{kl}\partial_j\partial_l\eta\\
\label{finalvar}
&=&(2\pi)^{-3/2}\eta+(3\pi\sqrt{2\pi})^{-1}\,\nabla^2\eta,
\end{eqnarray}
where the calculation of equation (\ref{finalvar}) is done in \ref{average}. Utilizing once more the wave equation $\nabla^2J_0=-J_0$ we obtain finally the large distance behaviour of the correlation function
\begin{equation}
\label{corrbridge}
\left<\chi(\bi{r})\chi(0)\right>=\frac{1}{72\pi^3}J_0(r)+\Or(J_0^3(r))\sim \cos(2\pi r) /\sqrt{r}.
\end{equation}
The correlations of $\chi$ are in fact long ranged. The prefactor, however, is extremely small, if compared with a naive guess for the $\chi$ correlation function, 
namely
\begin{equation}
\fl\left<\sign(\phi(\bi{r}))\sign(\phi(0))\right>\times\left(\textrm{density of saddle points}\right)^2=
\frac{2}{\pi}J_0(r) \times \frac{4}{3}+\ldots
\end{equation}
The prefactor of the naive approximation is $192\pi^2\approx1895$ times larger than the
prefactor of the exact asymptotic result. The numerical smallness of the $\chi$ correlations is the key to an understanding of the apparent agreement of the results of \cite{Bog02} with the real random waves \cite{Blu02}. One of the two following scenarios might be realized: Either the long-ranged correlations are irrelevant due to their oscillatory character and the system belongs to the short-range percolation universality class. Or, the behaviour of the system eventually crosses over to a behaviour, which is different from short-range percolation. 
Due to the smallness of (\ref{corrbridge}), the crossover will take place at very large systems (deep in the semiclassical regime)---for small systems, one will see percolative behaviour in agreement with \cite{Bog02}.


\section{Counting nodal domains with the help of Potts-spins}
In this section we study the percolative structure of the nodal domains of random waves. To be more specific, we attempt to calculate the moments of the number of nodal
 domains. As a byproduct we will obtain the probability, that two points at a distance $r$ belong to the same nodal domain.
To achieve this we make use of a Potts-model ``on top'' of the nodal structure. For regularization purposes, the nodal domains are placed on a quadratic lattice with a lattice parameter $a\ll 2\pi/k$. To each node $\bi{r}_i$ (where $i$ enumerates the nodes) of the lattice we assign beside the sign of the random wave $\sigma_i= \sign(\phi(\bi{r}_i))$ a Potts spin $s_i$ which takes values $1\ldots q$, where $q$ is a given natural number. We denote the pairs of adjacent nodes by $<i,j>$, i.e. $\sum_{<i,j>}$ is the sum over all bonds of the square lattice. We impose now a crucial restriction on the Potts-spins: All spins $s_i$, which belong to the same nodal domain, must have the same value: $s_i=s_j$ for all $i,j$ of the same nodal domain. Spins from different domains might have different values. The constraint is implemented in a local manner: The product over all lattice bonds
\begin{equation}
\label{constr}
\fl \prod_{<i,j>}\left(\frac{1-\sigma_i\sigma_j}{2}+\frac{1+\sigma_i\sigma_j}{2}\delta_{s_i,s_j}\right)=
\prod_{<i,j>}\left(1-\frac{1+\sigma_i\sigma_j}{2}(1-\delta_{s_i,s_j})\right)
\end{equation}
is one, if and only if all factors are equal to one. A certain factor is one, if either $\sigma_i\neq\sigma_j$ ($i,j$ belong to different domains) or if $i,j$ are in the same domain 
$\sigma_i=\sigma_j$ and the corresponding Potts spins are equal $\delta_{s_i,s_j}=1$.
Let us now sum over all possible spin configurations $\{s_i\}$. For each domain, the spins acquire the $q$ different values $1\ldots q$, i.e. each nodal domain contributes a factor of $q$ to the sum
\begin{equation}
\label{number}
\fl\mathcal{Z}[\{\sigma_i\}]=\sum_{\{s_i\}}\prod_{<i,j>}\left(1-\frac{1+\sigma_i\sigma_j}{2}(1-\delta_{s_i,s_j})\right)
=q^C
\end{equation}where $C$ is the number of nodal domains.

Now we perform the average $\left<\ldots\right>_\phi$ over the Gaussian random field $\phi$ with correlation function (\ref{corr}) and obtain the partition function
\begin{equation}
\label{expo}
\fl\mathcal{Z}=\sum_{\{s_i\}}\left<\exp\left(-\beta\sum_{<i,j>}\frac{1+\sigma_i\sigma_j}{2}(1-\delta_{s_i,s_j})\right)\right>_\phi\stackrel{\beta\rightarrow\infty}{\longrightarrow}
\left<q^C\right>_\phi\end{equation}
where we have in addition exponentiated the statistical weight (\ref{constr}). Equation (\ref{expo}) might be seen as a kind of (annealed) diluted Potts model similar to the one presented in \cite{Wu81,Nie82} with broken bonds along the nodal lines. 
Now we take the somewhat formal limit \cite{For72} $q\rightarrow 1$ and find for the mean number of nodal domains (the higher moments can be obtained analogously)
\begin{equation}
\left<C\right>_\phi=\left.\frac{\partial\mathcal{Z}}{\partial q}\right|_{q=1}\textrm{ for }\beta\rightarrow\infty.
\end{equation}
The connectivity of the domains can also be obtained with the help of the above partition function. Spins on different domains are independent, i.e. the correlation of the order parameter \begin{equation}
\psi_i=\delta_{s_i,1}-1/q
\end{equation}
vanishes:
\begin{equation}
\sum_{s_k=1\ldots q}\sum_{s_l}(\delta_{s_k,1}-1/q)(\delta_{s_l,1}-1/q)=0.
\end{equation}
If both spins are constrained to be equal, the correlation between $\psi_k$ and $\psi_l$ is in fact nonzero
\begin{equation}
\sum_s (\delta_{s,1}-1/q)^2=1-1/q.
\end{equation}
The correlation function reads now
\begin{eqnarray}
\lefteqn{\sum_{\{s_i\}}\psi_k\psi_l\prod_{<i,j>}\left(1-\frac{1+\sigma_i\sigma_j}{2}(1-\delta_{s_i,s_j})\right)}\nonumber\\
&=&q^{C}\times\left\{
\begin{array}{ll}0&\textrm{if }k,l\textrm{ are disconnected}\\
1-1/q&\textrm{if }k,l\textrm{ are connected}\end{array}\right..
\end{eqnarray}
Now we take the limit $q\rightarrow 1$
\begin{equation}
\left.\frac{\partial}{\partial q}\right|_{q=1}\left(q^{C}-q^{C-1}\right)=1,
\end{equation}perform the average $\left<\ldots\right>_\phi$ over the Gaussian random field $\phi$, and obtain finally the probability $p_{kl}$ that the points $k$ and $l$ are connected
\begin{equation}
\label{connect}
p_{kl}=\left.\frac{\partial}{\partial q}\right|_{q=1}\left<\sum_{\{s_i\}}\psi_k\psi_l\prod_{<i,j>}\left(1-\frac{1+\sigma_i\sigma_j}{2}(1-\delta_{s_i,s_j})\right)\right>_\phi
\end{equation}
The model (\ref{number}) is obviously a direct descendant of the Potts model, interpreted as the random-cluster model \cite{For72}. In the usual Potts model
\begin{equation}
\mathcal{Z}=\sum_{\{s_i\}}\prod_{<i,j>}\left(1+v\delta_{s_i,s_j}\right)
\end{equation}
the two processes of cluster generation and counting are interwoven---the expansion of the product generates the different clusters and the summation over the Potts-spins counts the number of clusters as in equation (\ref{number}). In our formula, we average over the random waves to generate the nodal domains, and average separately over the auxiliary Potts-spins to count the domains.

\section{Conclusion}
We could show, that the avoided intersections of Gaussian random waves show long-ranged correlations, which are not taken into account in the simple percolation model \cite{Bog02}.
It is therefore not evident, whether both models belong to the same universality class, i.e. share the same critical exponents.

In order to investigate the percolative structure of the nodal domains of random waves, we have proposed a model, which utilizes Potts-spins to count the number of domains. One might easily find variants of this model, e.g. by placing the auxiliary spins on the dual lattice in order to count the nodal lines or by using a ``height-variable'' as in \cite{Car94} to obtain the area distribution of the nodal domains.
All models suffer from the fact, that the auxiliary ``spins'' have infinite couplings ($\beta\rightarrow\infty$ in equation (\ref{expo})). Otherwise the interaction range of the auxiliary variables would be finite, making it impossible to decide whether distant regions are connected or not. Unfortunately, the infinite couplings complicate calculations and approximations considerably. The evaluation of the counting formula is therefore left to future investigations. A helpful first step would be the formulation of a continuous field theory corresponding to equation (\ref{expo}). Then, the relevance (or irrelevance) of the long-ranged correlations (\ref{corrbridge}) would become manifest.

\ack
It is a pleasure to thank U Smilansky for many enlightening discussions and for a critical reading of the manuscript. I acknowledge the hospitality of the Complex systems group at the Weizmann institute, where some of this work was done. 
I would like to thank A G Monastra and H K Janssen for useful discussions.
This work was supported by the Deutsche Forschungsgemeinschaft
under SFB 237, by the Minerva Center for non-linear Physics at the Weizmann Institute, and the Israel Science Foundation.

\appendix

\section{Calculation of the averages in equation (\ref{varrhoj})}

\label{average}
At first we calculate
\begin{eqnarray}
\fl-\left<\sign(\phi)\theta(-\det\partial\partial\phi)\partial_i\partial_k\phi\right>
&=&-(\delta_{ik}/2)\left<\sign(\phi)\nabla^2\phi\,\theta\left((\partial_x\partial_y\phi)^2-\partial_x\partial_x\phi\partial_y\partial_y\phi\right)\right>\nonumber\\
&=&(\delta_{ik}/2)\left<|\phi|\,\theta\left((\partial_x\partial_y\phi)^2-\partial_x\partial_x\phi\partial_y\partial_y\phi\right)\right>
\end{eqnarray}
due to the $O(2)$ symmetry and due to the fact, that $\phi$ obeys the wave equation $-\nabla^2\phi=\phi$. We introduce the Gaussian variables $X=\sqrt{8}\partial_x\partial_y\phi$, $Y=\sqrt{2}(\partial_x\partial_x-\partial_y\partial_y)\phi$ and $Z=(\partial_x\partial_x+\partial_y\partial_y)\phi=-\phi$, which are mutually independent random variables with
second moments $\left<X^2\right>=\left<Y^2\right>=\left<Z^2\right>=1$.
Then $-\det\partial\partial\phi=(X^2+Y^2-2Z^2)/8$ and
\begin{eqnarray}
\fl\lefteqn{
-\left<\sign(\phi)\theta(-\det\partial\partial\phi)\partial_i\partial_k\phi\right>}\nonumber\\
&=&(\delta_{ik}/2)\left<|Z|\,\theta(X^2+Y^2-2Z^2)\right>\nonumber\\
&=&\frac{\delta_{ik}}{2\sqrt{2\pi}}\int\rho^2\rmd\rho\sin(\vartheta)\rmd\vartheta\,\rho |\cos(\vartheta)|\,\theta\left(\sin^2(\vartheta)-2\cos^2(\vartheta)\right)\exp(-\rho^2/2)\nonumber\\
&=&\frac{\delta_{ik}}{\sqrt{2\pi}}\int_0^\pi\rmd\vartheta\sin(\vartheta)|\cos(\vartheta)|\,\theta\left(1-3\cos^2(\vartheta)\right)\nonumber\\
&=&\frac{\delta_{ik}}{\sqrt{2\pi}}\int_{-1}^1\rmd t\,|t|\,\theta\left(1-3t^2\right)=\frac{\delta_{ik}}{3\sqrt{2\pi}},
\end{eqnarray}
where we have introduced spherical coordinates $X=\rho\sin\vartheta\cos\varphi$, $Y=\rho\sin\vartheta\sin\varphi$, and $Z=\rho\cos\vartheta$. 
The other expectation value is
\begin{eqnarray}
2\left<\delta(\phi)F(-\det\partial\partial\phi)\right>&=&2\left<\delta(Z)F\left((X^2+Y^2-2Z^2)/8\right)\right>\nonumber\\
&=&\left<\delta(Z)\right>\left<(X^2+Y^2)\right>/4=\frac{1}{2\sqrt{2\pi}}.
\end{eqnarray}
With $\left<\delta^2(\nabla \phi)\right>=1/\pi$ we obtain finally
\begin{equation}
\delta_\eta\left<\chi(\bi{r})\right>=(2\pi)^{-3/2}\eta+(3\pi\sqrt{2\pi})^{-1}\,\nabla^2\eta.
\end{equation}


\section*{References}

\begin{thebibliography}{10}

\bibitem{Blu02}
Galya Blum, Sven Gnutzmann, and Uzy Smilansky.
\newblock Nodal domains statistics: A criterion for quantum chaos.
\newblock {\em Phys. Rev. Lett.}, 88:114101, 2002.

\bibitem{Str79}
Richard~M. Stratt, Nicholas~C. Handy, and William~H. Miller.
\newblock On the quantum mechanical implications of classical ergodicity.
\newblock {\em J. Chem. Phys.}, 71:3311--3322, 1979.

\bibitem{Sim96}
F.~Simmel and M.~Eckert.
\newblock Statistical measures for eigenfunctions of nonseparable quantum
  billiard systems.
\newblock {\em Physica D}, 97:517--530, 1996.

\bibitem{Ber77}
M.~V. Berry.
\newblock Regular and irregular semiclassical wavefunctions.
\newblock {\em J. Phys. A: Math. Gen.}, 10:2083--2091, 1977.

\bibitem{Pec72}
Philip Pechukas.
\newblock Semiclassical approximation of multidimensional bound states.
\newblock {\em J. Chem. Phys.}, 57:5577--5594, 1972.

\bibitem{Mon02}
Alejandro~G. Monastra, Uzy Smilansky, and Sven Gnutzmann.
\newblock Avoided intersections of nodal lines.
\newblock arXiv/nlin.CD/0212006, 2002.

\bibitem{Bog02}
E.~Bogomolny and C.~Schmit.
\newblock Percolation model for nodal domains of chaotic wave functions.
\newblock {\em Phys. Rev. Lett.}, 88:114102, 2002.

\bibitem{Wu81}
F~Y Wu.
\newblock Dilute potts model, duality and site-bond percolation.
\newblock {\em J. Phys. A: Math. Gen.}, 14:L39--L44, 1981.

\bibitem{Nie82}
B~Nienhuis.
\newblock Analytical calculation of two leading exponents of the dilute potts
  model.
\newblock {\em J. Phys. A: Math. Gen.}, 15:199--213, 1982.

\bibitem{For72}
C.~M. Fortuin and P.~W. Kasteleyn.
\newblock Random cluster model .1. introduction and relation to other models.
\newblock {\em Physica}, 57:536, 1972.

\bibitem{Car94}
John Cardy.
\newblock Mean area of self-avoiding loops.
\newblock {\em Phys. Rev. Lett.}, 72:1580--1583, 1994.

\end{thebibliography}
\end{document}